\documentclass[12pt]{article}
\usepackage{graphicx}
\usepackage{cite}
\usepackage{color}
\newcommand{\bm}[1]{\mbox{\boldmath $#1$}}
\newcommand{\fnd}[2]{\frac{\textstyle #1}{\textstyle #2}}
\newcommand{\x}[1]{{\textstyle #1}}
\newcommand{\abs}[1]{\left| #1\right|}
\newcommand{\fndrs}[4]{\fnd{\raisebox{#1}{$#2$}}{\raisebox{#3}{$#4$}}}
\newcommand{\xrm}[1]{{\textstyle \mbox{\rm #1}}}
\newcommand{\dissum}[2]{\displaystyle \sum_{#1}^{#2}}
\newcommand{\Real}[1]{\Re {\it e}\left(#1 \right)}
\newcommand{\Imag}[1]{\Im {\it m}(#1 )}
\begin{document}
\title{Pion-pion scattering near threshold in the Resonance-Spectrum Expansion}
\author{
Eef van Beveren\\
{\normalsize\it Centro de F\'{\i}sica Te\'{o}rica,
Departamento de F\'{\i}sica, Universidade de Coimbra}\\
{\normalsize\it P-3004-516 Coimbra, Portugal}\\
{\small http://cft.fis.uc.pt/eef}\\ [.3cm]
\and
George Rupp\\
{\normalsize\it Centro de F\'{\i}sica das Interac\c{c}\~{o}es Fundamentais,
Instituto Superior T\'{e}cnico}\\
{\normalsize\it Universidade T\'{e}cnica de Lisboa, Edif\'{\i}cio Ci\^{e}ncia,
P-1049-001 Lisboa, Portugal}\\
{\small george@ajax.ist.utl.pt}\\ [.3cm]
{\small PACS number(s): 14.40.-n, 14.40.Cs, 14.40.Ev, 14.40.Lb}
}

\maketitle

\begin{abstract}
We study the properties of the Resonance-Spectrum Expansion near threshold
in $I=0$ $S$-wave $\pi\pi$ scattering. The real part of the amplitude, when
extrapolated from above threshold to below threshold, is found to vanish at
a positive non-zero value of the total invariant mass of the system,
in agreement with predictions from perturbative chiral models.
In our exact analytic expression, the total amplitude vanishes identically
at zero invariant mass.
\end{abstract}

\section{Introduction}
The Resonance-Spectrum Expansion (RSE) has been developed \cite{IJTPGTNO11p179}
for the description of meson-meson scattering resonances and bound states, in
a non-perturbative approach that aims at unquenching the $q\bar{q}$ confinement
spectrum. It consists of a simple analytic expression which can be
straightfrowardly applied to all possible non-exotic systems of two mesons.
The RSE goes beyond simple spectroscopy, since it describes the scattering
amplitude, not only at a resonance, but also for energies where no resonance
exists. In contrast to models which have to rely upon numerical methods of
solution, the RSE has the additional advantage that the pole structure of its
scattering amplitude can be studied in great detail, owing to its closed
analytic form. The expression for the amplitude can even be analytically
continued, in an exact manner, to below the lowest threshold, where bound
states show up as poles on the real energy axis. The RSE easily handles many
coupled meson-meson channels, or coupled systems with different internal
flavours. As such, it is an ideal expression for the study of scattering theory
in general, i.e., the study of resonance structures and their relation to
some of the many $S$-matrix singularities, as well as the concept of Riemann
sheets and analytic continuation anywhere in the complex energy plane.
Moreover, it is particularly powerful in examining the properties of
scattering amplitudes near the lowest threshold, since there it can effectively
be reduced to a one-channel case, with just two Riemann sheets.
In the present paper, we shall study isoscalar $S$-wave $\pi\pi$ scattering
near threshold, thereby ignoring the small mass difference between neutral
and charged pion pairs.

The basic ingredients of the RSE are confinement and quark-pair creation.
In its lowest-order approximation, a permanently confined quark-antiquark
system is assumed, having a spectrum with an infinite number of bound states,
related to the details of the confining force. We shall denote the energy
levels of this confinement spectrum by
$E_{n}$ ($n=0$, 1, 2, $\dots$).
Here , we assume that the confinement spectrum is given by
\begin{equation}
E_{n}\; =\; E_{0}\, +\, 2n\omega
\;\;\; ,
\label{t00HO}
\end{equation}
which choice is anyhow rather immaterial for the purpose of our present study.
The parameter $E_{0}$ represents, to lowest order, the ground-state mass of
the quark-antiquark system, which is related to the effective flavour masses
of the system. The strength of the confinement force is parametrised
by $\omega$ and gauged by the level splittings of the system.
In experiment one cannot directly measure the quantities
$E_{0}$ and $\omega$, because of the large higher-order contributions.
Consequently, the lowest-order system is purely hypothetical.
Nevertheless, one can obtain some order-of-magnitude insight
by examining mesonic spectra with more than one experimentally known
recurrency.

From the $J^{PC}=1^{--}$ charmonium and bottomonium states, one may conclude
that the average level splitting is of the order of 380 MeV, leading to
$\omega =0.19$ MeV, independent of flavour.
The latter property is compatible with the flavour blindness of QCD,
confirmed by experiment \cite{PRD59p012002}.
Indeed, the level splittings of the positive-parity $f_{2}$ mesons
seem to confirm that flavour independence can be extended
to light quarks \cite{HEPPH0610199}.
From the ground states of the recurrencies one may then extract
the order of magnitude of the effective quark masses,
e.g.\ $2m_{c}=m(J/\psi )=3.1$ GeV (in the RSE \cite{PRD27p1527}
we find for twice the effective charm mass the value 3.124 GeV),
or $2m_{u/d}=m(\rho )=0.77$ GeV (0.812 GeV in the RSE).
For the choice (\ref{t00HO}) of confinement force,
we determine $E_{0} = m_{q} + m_{\bar{q}} + (1.5 + \ell ) \omega$.
Once the effective flavour masses and $\omega$ are fixed \cite{PRD27p1527},
we may describe other systems, like scalar mesons and mixed flavours
\cite{ZPC30p615,PRL91p012003,HEPPH0312078,MPLA19p1949,PRL97p202001}.

Through quark-pair creation the $q\bar{q}$ system is coupled
to those two-meson systems which are allowed by quantum numbers.
In principle, many different two-meson channels can couple to
one specific quark-antiquark system. Here, since we study the properties
of the channel lowest in mass, we will strip the RSE of all other possible
two-meson channels, thereby assuming that their influence far below their
respective thresholds will be negligible.
Via consecutive quark-pair creation and annihilation,
a $q\bar{q}$ pair may also couple to pairs of different flavour,
for instance $u\bar{u}(d\bar{d})\leftrightarrow s\bar{s}$.
Here, as we study pion-pion scattering near threshold,
we shall assume that the coupling of a light pair of flavours
to strange-antistrange can be ignored.

The intensity of quark-pair creation is in the RSE parametrised by
the flavour-independent parameter $\lambda$. In principle, it has to be
adjusted to the data.  However, one may get an idea of the right order of
magnitude by the following reasoning. For small values of $\lambda$, one may
determine the width of the ground-state resonance in the one-channel case
(pion-pion here) by \cite{AP105p318}
\begin{equation}
\Gamma\;\approx\;
\lambda^{2}\, E_{0}\,
\fndrs{8pt}
{\sin^{2}\left(\fndrs{2pt}{a}{-3pt}{2}\sqrt{E_{0}^{2}-4m^{2}_{\pi}}\,\right)}
{-5pt}
{\fndrs{2pt}{a}{-3pt}{2}\sqrt{E_{0}^{2}-4m^{2}_{\pi}}}
\;\;\; ,
\label{RSEwidth}
\end{equation}
where $a$ represents the average distance at which light quark pairs are
created, and which can also be defined in a flavour-independent fashion
\cite{PRD21p772}. For light quarks, $a$ is about 0.6 fm, as we will see
later on. If we take, for example, the $f_{0}$(1370) resonance width of
0.2--0.5 GeV, then we obtain $\lambda\approx$ 0.60--0.95.
This is of the order of 1, which would not allow the approximation
(\ref{RSEwidth}). Nevertheless, we actually employ here a value for $\lambda$
which is of the same order of magnitude as our estimate
(see caption of Fig.~\ref{t00Figures}).

\section{The RSE amplitude for the \bm{I=0} \bm{\pi\pi} isoscalar \bm{S}-wave}

The RSE amplitude suitable for our purposes results from the ladder sum in
quark-pair creation \cite{IJTPGTNO11p179},
and has for $I=0$ $S$-wave $\pi\pi$ scattering the form
\begin{equation}
t_{0}^{0}(E)\; =\;\fnd{E}{2k}\, T_{0}^{0}(E)\; =\;
\fndrs{12pt}
{a\lambda^{2}\, E\mu\,
\fndrs{5pt}{\sin^{2}(ka)}{-5pt}{(ka)^{2}}
\dissum{n=0}{\infty}\,\fndrs{3pt}{(n+1)\, 4^{-n}}{-3pt}{E_{n}-E}\,
}
{-15pt}
{1-2\lambda^{2}\,\mu\,
\fndrs{5pt}{\sin (ka)}{-5pt}{ka}\, e^\x{ika}\,
\dissum{n=0}{\infty}\,\fndrs{3pt}{(n+1)\, 4^{-n}}{-3pt}{E_{n}-E}
}
\;\;\; ,
\label{t00RSE}
\end{equation}
where $E_{n}$ is defined in Eq.~(\ref{t00HO}), and where
\begin{equation}
k\; =\; k(E)\; =\;\frac{1}{2}\sqrt{E^{2}-4m_{\pi}^{2}}
\;\;\;\;\xrm{and}\;\;\;\;
\mu\; =\;\mu(E)\; =\;\frac{1}{4}E
\;\;\; .
\label{kandmuinE}
\end{equation}
The factor $(n+1)\, 4^{-n}$ is a remnant \cite{ZPC21p291}
of the quark-antiquark distributions associated
with the confinement spectrum (\ref{t00HO}).
The amplitude (\ref{t00RSE}) satisfies the unitarity condition
$\abs{1+2iT_{0}^{0}(E)}=1$ for all energies $E>2m_{\pi}$,
as can be easily verified.
\begin{figure}[htbp]
\begin{center}
\begin{tabular}{|c|c|}
\hline & \\ [-7pt]
\resizebox{0.47\textwidth}{!}{\includegraphics{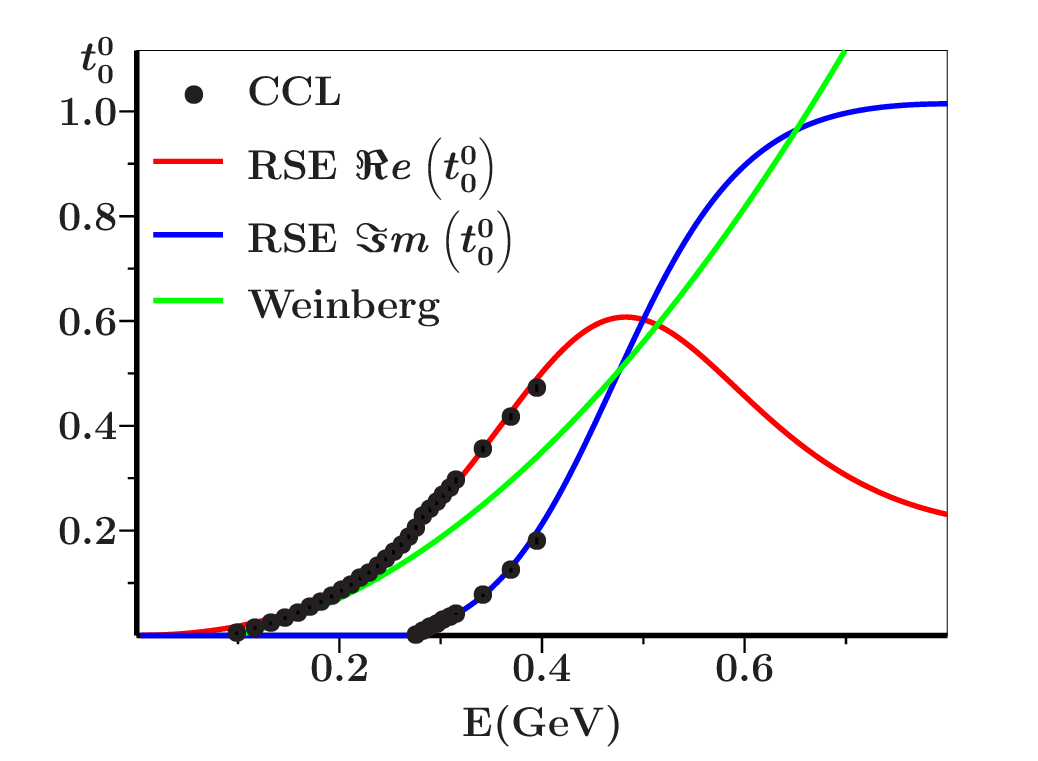}} &
\resizebox{0.47\textwidth}{!}{\includegraphics{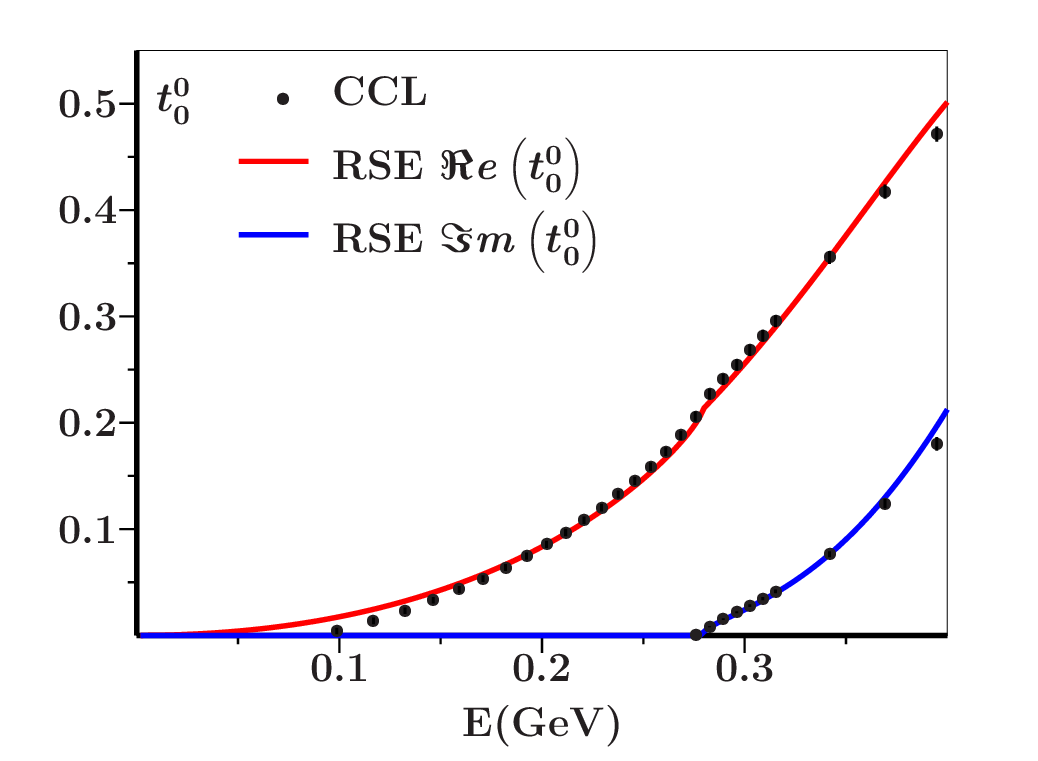}}\\ [3pt]
\hline
\end{tabular}
\end{center}
\caption[]{\small $\Real{t_{0}^{0}}$ (red line) and $\Imag{t_{0}^{0}}$
(blue line) for the RSE of Eq.~\ref{t00RSE}, compared with
Weinberg's expression of Eq.~\ref{t00Weinberg} (green line),
for $f_{\pi}=92.4$ MeV, and the results of Caprini, Colangelo and Leutwyler
(CCL) \cite{PRL96p132001} (black dots).
The RSE parameters are $\lambda =  1.29$, $a =  2.90$ GeV$^{-1}$,
$E_{0} =  1.30$ GeV and $\omega =  0.19$ GeV.}
\label{t00Figures}
\end{figure}

In Fig.~\ref{t00Figures} we compare the RSE amplitude
with the amplitude of Weinberg \cite{PRL17p616}
and the dispersion-relation result of Caprini, Colangelo and Leutwyler
\cite{PRL96p132001}.
Weinberg's relation is given by
\begin{equation}
t_{0}^{0}(E)\; =\;\fndrs{5pt}{E^{2}-\frac{1}{2}m_{\pi}^{2}}
{-5pt}{16\pi f_{\pi}^{2}}
\;\;\; .
\label{t00Weinberg}
\end{equation}
We do not distinguish here between neutral and charged pions,
and take the pion mass equal to the one of the charged pions.
At threshold we find for the RSE
\begin{equation}
t_{0}^{0}\left( E=2m_{\pi}\right)\;
=\;\Real{t_{0}^{0}}\left( E=2m_{\pi}\right)\;
=\; 0.212
\;\;\; ,
\label{t00atthreshold}
\end{equation}
which may be compared to data at $0.220\pm 0.005$
\cite{NPB603p125}.

In the lefthand side picture of Fig.~\ref{t00Figures}
one observes from the behaviour of both
$\Real{t_{0}^{0}}$ and $\Imag{t_{0}^{0}}$ that the RSE clearly describes
a resonant structure, i.e., the $f_{0}$(600) alias $\sigma$ meson.
However, for energies above 600 MeV,
the RSE prediction does not follow the data.
The main reason is the absence of a coupling to $s\bar{s}$ in the $q\bar{q}$
sector, as well as the $K\!\bar{K}$ channel (see e.g.\ Ref.~\cite{PLB641p265}).
However, for energies below 400 MeV
the agreement with the data is excellent.
Even below threshold, at about 280 MeV, the hypothetical data of
Ref.~\cite{PRL96p132001} are fairly well reproduced.
In fact, the RSE only deviates because it does not exactly reproduce
the so-called Adler zero \cite{PR137pB1022} for non-vanishing total invariant
mass $E$. This does not mean that above threshold, where the RSE scattering
amplitude does agree with the true data, the real part of $t_{0}^{0}$ cannot
be proportional to something of a form similar to Weinberg's expression
(\ref{t00Weinberg}). It only means that below threshold the analytic form of
$\Real{t_{0}^{0}}=t_0^0$ may be slightly different from what is predicted in
Refs.~\cite{PRL17p616,PR137pB1022}, for very small, unphysical values of $E$.

\begin{figure}[htbp]
\begin{center}
\resizebox{0.47\textwidth}{!}{\includegraphics{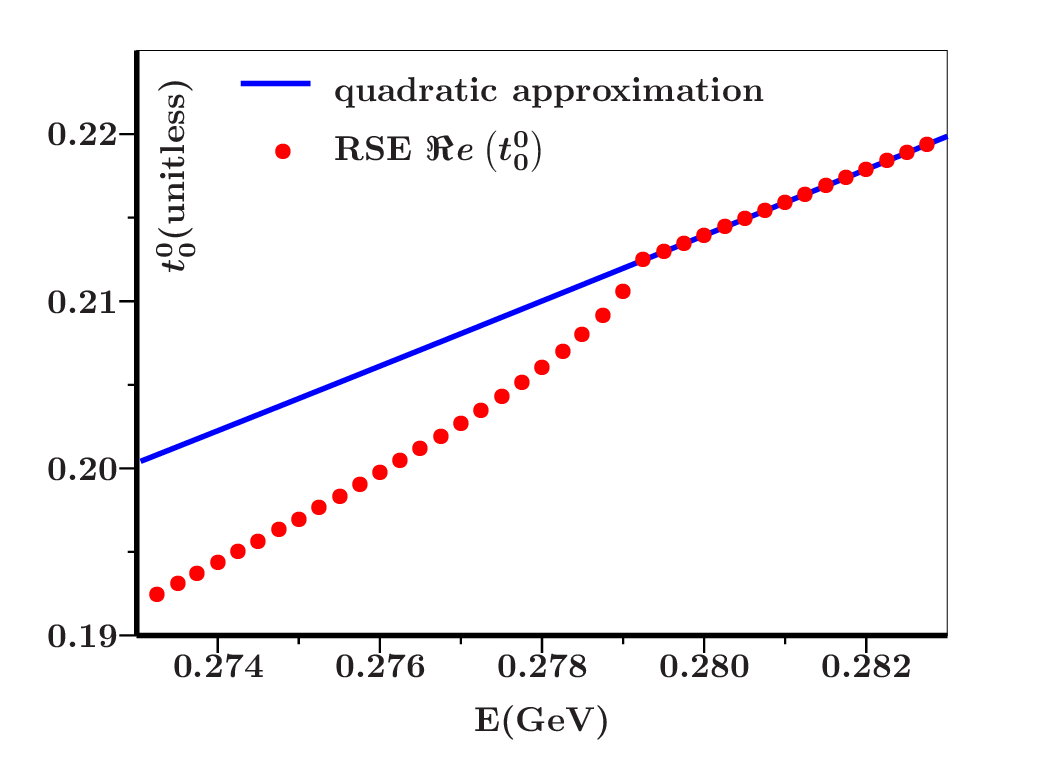}}\\ [3pt]
\end{center}
\caption[]{\small $\Real{t_{0}^{0}}$ and its quadratic approximation
(Eqs.~\ref{t00thrapprox} and \ref{t00thrapproxAB})
for the resonance-spectrum expansion (RSE) (Eq.~\ref{t00RSE}),
(respectively red dots and blue line). Threshold is at 279.14 MeV
for charged pion pairs.}
\label{t00FigThreshold}
\end{figure}

In order to study this in more detail,
we expand formula (\ref{t00RSE}) near threshold ($E>2m_{\pi}$):
\begin{eqnarray}
\lefteqn{\Real{t_{0}^{0}(E)}\;\approx\;
\fnd{a}{4m_{\pi}}\,
\fndrs{5pt}{
\left( 1-\frac{1}{3}a^{2}m^{2}_{\pi}\right) S_{1}+S_{2}
-\frac{3}{2}S_{1}^{2}-S_{1}S_{2}
+\left(\frac{1}{2}-\frac{2}{3}a^{2}m^{2}_{\pi}\right) S_{1}^{3}}
{-5pt}{\left( 1-S_{1}\right)^{3}}\,\times}
\nonumber\\ [10pt] & & \times\,
\left\{\, E^{2}\, +\,
4m_{\pi}^{2}\,\fndrs{5pt}
{\frac{1}{3}a^{2}m^{2}_{\pi}S_{1}-S_{2}
-\frac{1}{2}S_{1}^{2}+S_{1}S_{2}
+\left(\frac{1}{2}+\frac{2}{3}a^{2}m^{2}_{\pi}\right) S_{1}^{3}}
{-5pt}
{\left( 1-\frac{1}{3}a^{2}m^{2}_{\pi}\right) S_{1}+S_{2}
-\frac{3}{2}S_{1}^{2}-S_{1}S_{2}
+\left(\frac{1}{2}-\frac{2}{3}a^{2}m^{2}_{\pi}\right) S_{1}^{3}}
\right\}
\;\;\; ,
\label{t00thrapprox}
\end{eqnarray}
where we have defined
\begin{equation}
S_{\alpha}\, =\,\lambda^{2}\dissum{n=0}{\infty}
\fndrs{0pt}{m_{\pi}^{\alpha}(n+1)\, 4^{-n}}
{-5pt}{\left( E_{0}+2n\omega-2m_{\pi}\right)^{\alpha}}
\;\;\;\; (\alpha =1,2)
\;\;\; .
\label{Salphadef}
\end{equation}
Insertion of the model parameters (see caption of Fig.~\ref{t00Figures})
yields
\begin{equation}
\Real{t_{0}^{0}(E)}\;\approx\;
\fndrs{0pt}{E^{2}\, -\, 0.90\, m^{2}_{\pi}}
{-5pt}{16\pi\,\left( 75.2\;\;\xrm{MeV}\right)^{2}}
\;\;\; .
\label{t00thrapproxAB}
\end{equation}
Consequently, we find that the apparent ``Adler zero'' in the RSE, as seen
from above threshold, is almost twice the value that follows
from Eq.~\ref{t00Weinberg}. Anyhow, the behaviour of $\Real{t_{0}^{0}}$ below
threshold cannot not be a simple continuation of the form above threshold,
since the derivative of $\Real{t_{0}^{0}}$ at threshold is discontinuous,
as is shown in Fig.~\ref{t00FigThreshold}.
The crucial point is that one cannot analytically continue the real part of
the amplitude, for the trivial reason that the real part of an analytic
function is not analytic.
Hence, where $t_0^0\equiv\Real{t_{0}^{0}}$ (for $E<2m_\pi$)
precisely has its zero
is rather irrelevant for the behaviour of $\Real{t_{0}^{0}}$ for $E>2m_\pi$.
For example, the twice-subtracted dispersion relations
of Ref.~\cite{PRL96p132001}
yield a zero at $(0.41\pm 0.06)\, m^{2}_{\pi}$
for the amplitude.
However, when the real part of the same amplitude
is extrapolated from above threshold to below threshold,
according to an approximation of the form (\ref{t00thrapproxAB}),
then one finds a zero at $(0.80\pm 0.10)\, m^{2}_{\pi}$.

Alternatively, we may consider the modulus of the amplitude
(\ref{t00RSE}), for which the derivative at threshold is continuous,
and which behaves near threshold according to
\begin{equation}
\abs{t_{0}^{0}(E)}\;\approx\;
\fndrs{0pt}{E^{2}\, -\, 0.95\, m^{2}_{\pi}}
{-5pt}{16\pi\,\left( 74.6\;\;\xrm{MeV}\right)^{2}}
\;\;\;\; (E>2m_{\pi}).
\label{tooRSEmodulus}
\end{equation}

This is probably where Hanhart in Ref.~\cite{HEPPH0609136} mixed up
the RSE with perturbative considerations, when claiming that
the RSE amplitude does not behave properly at threshold
because the amplitude vanishes for $E=0$.
But even in a perturbative approach to the RSE amplitude
(\ref{t00RSE}) for small $\lambda$, i.e.,
\begin{eqnarray}
\lefteqn{\left\{ t_{0}^{0}\right\}_\xrm{\scriptsize pert}(E)\; =\;
a\lambda^{2}\, E\mu\,
\fndrs{5pt}{\sin^{2}(ka)}{-5pt}{(ka)^{2}}
\dissum{n=0}{\infty}\,\fndrs{3pt}{(n+1)\, 4^{-n}}{-3pt}{E_{n}-E}}
\nonumber\\ [10pt] & & \approx\,
\fnd{a}{4m_{\pi}}\left\{
\left( 1-\frac{1}{3}a^{2}m^{2}_{\pi}\right) S_{1}+S_{2}\right\}
\left[ E^{2}-4m_{\pi}^{2}\fndrs{5pt}
{-\frac{1}{3}a^{2}m^{2}_{\pi}S_{1}+S_{2}}
{-5pt}
{\left( 1-\frac{1}{3}a^{2}m^{2}_{\pi}\right) S_{1}+S_{2}}
\right]
\;\;\; ,
\label{t00RSEpert}
\end{eqnarray}
which is real on the real-energy axis
and which clearly vanishes at $E=0$,
one obtains an extrapolated zero at
$0.25\, m^{2}_{\pi}$ for the RSE model parameters
(see caption Fig.~\ref{t00Figures}). Namely,
\begin{equation}
\left\{ t_{0}^{0}\right\}_\xrm{\scriptsize pert}(E)\;\approx\;
\fndrs{0pt}{E^{2}\, -\, 0.251\, m^{2}_{\pi}}
{-5pt}{16\pi\,\left( 102\;\;\xrm{MeV}\right)^{2}}
\;\;\;\; (E>2m_{\pi}).
\label{tooRSEpertAB}
\end{equation}
The Adler-Weinberg zero at $0.50\, m^{2}_{\pi}$
only holds in lowest order in the chiral expansion \cite{PRL17p616}.
Higher-order terms then move it to a different value
\cite{PRL96p132001}.
We observe here that the higher-order corrections are substantial,
by comparing the values obtained from the threshold behaviour
in the RSE
for the amplitude's Born term (Eq.~\ref{tooRSEpertAB})
and for the full amplitude,
either for its real part (Eq.~\ref{t00thrapproxAB}),
or alternatively for its modulus (Eq.~\ref{tooRSEmodulus}).

\section{Conclusions}

The predictive power of the RSE as an analytic method to unquench the
quark model has been demonstrated before, by interrelating an enormous variety
of non-exotic mesonic systems, such as the light scalar mesons
$f_{0}$(600), $f_{0}$(980), $K_{0}^{\ast}$(800), $a_{0}$(980) and the
corresponding $S$-wave $\pi\pi$, $K\pi$, $\eta\pi$ scattering
observables \cite{ZPC30p615,PLB641p265}, the scalars between 1.3 and
1.5 GeV \cite{ZPC30p615}, vector and pseudoscalar mesons \cite{PRD27p1527},
charmonium and bottomonium \cite{PRD21p772}, the $D_{s0}^{\ast}$(2317)
\cite{PRL91p012003}, and the $D_{sJ}^{\ast}$(2860) \cite{PRL97p202001}.
In the present Letter, we have shown that also at the $\pi\pi$ threshold
the RSE behaves as expected from more general considerations.

\section*{Acknowledgments}
This work was supported in part by the {\it Funda\c{c}\~{a}o para a
Ci\^{e}ncia e a Tecnologia} \/of the {\it Minist\'{e}rio da Ci\^{e}ncia,
Tecnologia e Ensino Superior} \/of Portugal, under contract
PDCT/FP/63907/2005.

\newcommand{\pubprt}[4]{#1 {\bf #2}, #3 (#4)}
\def\AP{Annals Phys.}
\def\IJTPGTNO{Int.\ J.\ Theor.\ Phys.\ Group Theor.\ Nonlin.\ Opt.}
\def\MPLA{Mod.\ Phys.\ Lett.\ A}
\def\NPB{Nucl.\ Phys.\ B}
\def\PLB{Phys.\ Lett.\ B}
\def\PR{Phys.\ Rev.}
\def\PRD{Phys.\ Rev.\ D}
\def\PRL{Phys.\ Rev.\ Lett.}
\def\ZPC{Z.\ Phys.\ C}

\end{document}